\documentclass[twocolumn]{emulateapj}

\newcommand{\expval}[1]{\left\langle #1 \right\rangle}

\shorttitle{Parameterization of Gamma-ray Angular Distribution}
\shortauthors{Karlsson \& Kamae}

\begin{document}


\title{Parameterization of the Angular Distribution of Gamma Rays Produced by p-p Interaction in Astronomical Environment}


\author{Niklas Karlsson\altaffilmark{1} and Tuneyoshi Kamae\altaffilmark{2,3}}
\affil{Stanford Linear Acceleration Center, Menlo Park, CA 94025}
\email{niklas@slac.stanford.edu}


\altaffiltext{1}{Visiting scientist from the Royal Institute of Technology, AlbaNova University
Center, SE-106 91 Stockholm, Sweden} 
\altaffiltext{2}{Also with the Kavli Institute for Particle Astrophysics and Cosmology, Stanford
University, Menlo Park, CA 94025}
\altaffiltext{3}{Electronic address: kamae@slac.stanford.edu}

\journalinfo{The Astrophysical Journal, accepted 2007 October 8}
\submitted{Submitted 2007 August 17; accepted 2007 October 8}


\begin{abstract}
We present the angular distribution of gamma rays produced by proton-proton interactions in
parameterized formulae to facilitate calculations in astrophysical environments. The
parameterization is derived from Monte Carlo simulations of the up-to-date proton-proton interaction
model by \citet{article:KamaeAbeKoi:2005} and its extension by \citet{article:Kamae_etal:2006}. This
model includes the logarithmically rising inelastic cross section, the diffraction dissociation
process and Feynman scaling violation. The extension adds two baryon resonance contributions: one
representing the $\Delta(1232)$ and the other representing multiple resonances around 1600
MeV/$c^{2}$. We demonstrate the use of the formulae by calculating the predicted gamma-ray spectrum
for two different cases: the first is a pencil beam of protons following a power law and the second
is a fanned proton jet with a Gaussian intensity profile impinging on the surrounding material. In
both cases we find that the predicted gamma-ray spectrum to be dependent on the viewing angle.
\end{abstract}





\section{Introduction}
Gamma-ray emission due to decays of neutral pions produced in proton-proton ($p$-$p$) interactions
has been predicted from the Galactic ridge, supernova remnants (SNRs), active galactic nucleus (AGN)
jets, and other astronomical sites \citep{book:Hayakawa:1969, article:Stecker:1970,
book:MurthyWolfendale:1986, book:Schonfelder:2001, book:Schlickeiser:2002, book:Aharonian:2004}. A
multitude of gamma-ray sources are already known today
\citep[see, e.g.,][]{article:Hartman_etal:1999} and with new gamma-ray observatories covering GeV to
TeV energies many more are expected to be found \citep{article:Aharonian_etal:2003,
article:Aharonian_etal:2004a, article:Aharonian_etal:2004b, article:Aharonian_etal:2005,
article:Ong:1998, article:Schroedter_etal:2005, book:Schonfelder:2001, inproc:Weekes:2003}. The GeV
energy window, in particular above about 10 GeV, has been poorly explored and the GLAST Large Area
Telescope (GLAST-LAT)\footnote{GLAST Large Area Telescope, http://www-glast.stanford.edu.} is
expected to provide high-statistics data in this window.

Interpretation of the observed gamma-ray spectra and identification of the involved interactions
require not only high-quality observational data, but also good knowledge of the contributing
production mechanisms. In the high-energy regime (sub-GeV to multi-TeV energies), the two dominant
processes for gamma-ray production are $p$-$p$ interactions and subsequent decays of neutral
pions, and inverse Compton (IC) up-scattering of low-energy photons on high-energy electrons.
Production of gamma rays through pion decay relies on acceleration of cosmic ray primaries, protons
or heavier nuclei, to high energies. 

The diffuse gamma-ray emission from the Galactic ridge was first detected by the OSO-3 and SAS-2
satellites and later the COS-B and EGRET instruments. This emission is interpreted as predominantly
due to decays of neutral pions produced in interactions between accelerated protons or heavier
nuclei with the interstellar medium \citep[ISM;][]{article:Stecker:1973, inbook:Stecker:1989,
article:Strong_etal:1978, article:Strong_etal:1982, article:Strong_etal:2000,
article:Strong_etal:2004, article:Strong_etal:2007, article:StephensBadhwar:1981,
article:Dermer:1986a, article:Hunter_etal:1997}. The gamma-ray flux and spectral shape measured by
EGRET \citep{article:Hunter_etal:1997} is considered as the key attestation of this interpretation.
There is also significant contribution to the diffuse gamma-ray emission from IC scatterings, in
particular in the Galactic ridge \citep{book:MurthyWolfendale:1986, article:Strong_etal:2000,
book:Schonfelder:2001}.

Several SNRs have been detected in TeV energies with ground based Air Cherenkov Telescopes (ACTs),
including the shell-type SNRs RX J1713.7-3946 and RX J0852.0-4622
\citep{article:Aharonian_etal:2004a, article:Aharonian_etal:2005}. Observations in the X-ray band
from both RX J1713.7-3946 \citep{article:Koyama_etal:1997, article:Slane_etal:1999,
article:UchiyamaAharonianTakahashi:2003} and RX J0852.0-4622 \citep{article:Tsunemi_etal:2000,
article:Iyudin_etal:2005} show a smooth, featureless spectrum indicating synchrotron X-ray emission
due to a population of TeV electrons. The same electron population might also produce high-energy
gamma rays through IC scatterings, but the measured gamma-ray fluxes and spectra in TeV
energies do not fully match the predicted ones
\citep[see, e.g., the analysis in][]{article:UchiyamaAharonianTakahashi:2003}, and it has been
suggested that there might also be a significant component due to hadronic interactions with the
surrounding ISM \citep{article:BerezhkoVolk:2000, article:Enomoto_etal:2002, book:Aharonian:2004,
article:Katagiri_etal:2005}. Hadronic models fit the very-high-energy gamma-ray spectrum assuming a
beam of accelerated protons \citep{inproc:Moskalenko_etal:2007}. The highest-energy cosmic rays
(CRs) escape the forward shock region of the SNR almost unidirectionally; the gamma-ray spectrum
becomes angular dependent.

High-energy gamma-ray emission from AGN jets is usually explained using leptonic models with
electrons accelerated to TeV energies. Support for this comes from the observed radio and X-ray
spectra which match those of synchrotron radiation from high-energy electron populations. The
apparent synchronization in the observed variability of X-ray and gamma-ray fluxes gives further
support for the leptonic modeling \citep{article:Ong:1998, book:Schonfelder:2001,
book:Schlickeiser:2002, book:Aharonian:2004}. There are observations of AGN jets where the leptonic
scenario faces difficulties. For these jets gamma-ray production through $p$-$p$ interactions has
been put forward as an alternative \citep{article:MuckeProtheroe:2001, article:Mucke_etal:2003,
article:BottcherReimer:2004}.

The giant radio galaxy M87 has recently been observed in TeV energies with the H.E.S.S.\ Cherenkov
telescopes \citep{inproc:Beilicke_etal:2005}. The jet is aligned about 30$^{\circ}$
\citep{article:BicknellBegelman:1996} relative to the line of sight and it has been well studied in
radio, optical and X-ray wavelengths. The central object is supposedly a super-massive black hole.
\citet{article:Stawarz_etal:2006} interpret the TeV emission as due to Componization of synchrotron
radiation from a flare in the nucleus. \citet{article:ReimerProtheroeDonea:2004}, on the other hand,
have suggested the Synchrotron-Proton Blazar (SPB) model as the production mechanism for the TeV
gamma-ray emission. The SPB model require protons to be accelerated to extremely high energies
making it less favorable. Another possibility is accelerated protons leaking out of the jet near the
central object, interacting with surrounding material and producing gamma-rays through the decay of
neutral pions.

In this paper the angular distribution of gamma rays produced in proton-proton interactions is
presented in parameterized formulae. These are derived from Monte Carlo simulations of the
up-to-date proton-proton interaction model by \citet{article:Kamae_etal:2006}. The angular
distribution is given relative to the incident proton direction. With this formalism, the gamma-ray
spectrum can be calculated for any given distribution of protons, including angular dependent ones.

Parameterization of the angular distributions of other stable secondary particles, i.e. electrons,
positrons and neutrinos, has been deferred due to observational limitations. When high statistics
neutrino data becomes available \citep{inproc:Halzen:2005} it may be worth while extending the
parameterization to include the angular distribution of neutrinos.

It is noted that \citet{article:KoersPeerWijers:2006} have taken another approach and parameterized
the energy and rapidity distributions of pions and kaons and from this they are able to derive the
spectrum and angular distribution of gamma rays. They consider proton energies above 1 TeV and
therefore their model is not suited for studies in the GLAST-LAT energy range. In this paper, proton
energies from the pion production threshold and the resonance region up to about $10^{5}$ GeV are
considered. High-precision data is expected from the GLAST-LAT in the GeV range which makes a
parameterization covering this range important 

\section{Proton-Proton Interaction Model}
For this work the proton-proton interaction model used to calculate the diffuse Galactic
gamma-ray flux \citep{article:KamaeAbeKoi:2005} and its extension \citep{article:Kamae_etal:2006} is
adopted. In an effort to predict the contribution to the Galactic diffuse emission from $\pi^{0}$
decays, \citet{article:KamaeAbeKoi:2005} found that past calculations \citep{article:Stecker:1970,
article:Stecker:1973, inbook:Stecker:1989, article:Strong_etal:1978, article:StephensBadhwar:1981,
article:Dermer:1986a, article:Dermer:1986b, article:Mori:1997} left out two important features of
the inelastic $p$-$p$ interaction, the diffraction dissociation process and the Feynman scaling
violation. It was also noted that past calculations assumed an energy independent inelastic $p$-$p$
cross section, about 24 mb for $T_{p}\gg 10$ GeV, in contradiction to recent experimental data where
a logarithmic increase of the cross section with the incident proton energy is evident
\citep{article:Hagiwara_etal:2002}. The predicted gamma-ray spectrum changed significantly in the
GeV energy range when the above features were included. The power-law index of the gamma-ray
spectrum is about 0.05 lower in absolute value than that of the incident proton spectrum and the
gamma-ray flux is increased significantly compared to the reference scaling model
\citep{article:KamaeAbeKoi:2005}; the increase is proton energy dependent, about 10-20\% in the
low-GeV range and about 50\% above a few 100 GeV. From here on, the model by
\citet{article:KamaeAbeKoi:2005} is referred to as model A.

\begin{figure}
\begin{center}
\scalebox{1.0}{\plotone{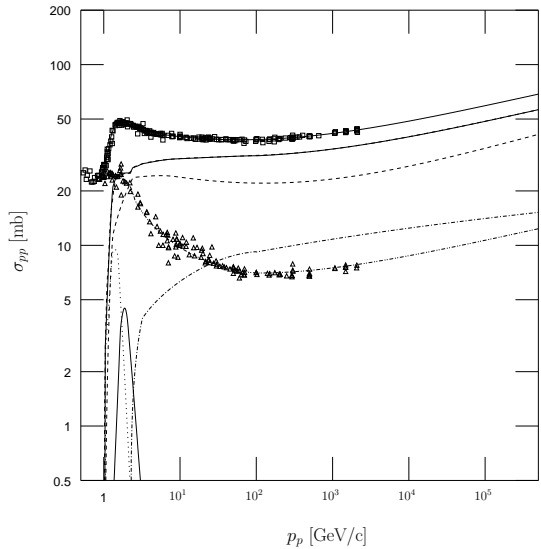}}
\end{center}
\caption{Experimental $p$-$p$ cross sections, as a function of proton momentum, and that of adjusted
model A: experimental total (squares), experimental elastic (triangles), total inelastic (thick
solid line), non-diffractive (dashed line), diffractive process (dot-dashed line), $\Delta$(1232)
(dotted line), and res(1600) (thin solid line). The total inelastic is the sum of the four
components. The thin solid and dot-dot-dashed lines running through the two experimental data sets
are eye-ball fits to the total and elastic cross sections, respectively. The functional forms are
those given by equations (1) through (4) in \citet{article:Kamae_etal:2006}.}
\label{fig:pp_cross_sections}
\end{figure}

Model A was primarily aimed at studying the diffuse emission from the Galactic ridge and thus
concerned proton kinetic energies well above 1 GeV. It is not accurate near the pion production
threshold \citep[see Figure 5 of][]{article:KamaeAbeKoi:2005}. To correct for this and improve the
accuracy for lower proton momenta, \citet{article:Kamae_etal:2006} adjusted model A by including
contributions from two baryon resonance excitation states, $\Delta(1232)$, representing the physical
$\Delta$ resonance with a mass of 1232 MeV/$c^{2}$, and res(1600), representing several resonances
with masses around 1600 MeV/$c^{2}$. The term ``baryon resonance'' refers to both nucleon resonances
with iso-spin 1/2 and $\Delta$ resonances with iso-spin 3/2. The $\Delta(1232)$ decays to a nucleon
(proton or neutron) and one pion
\citep[$\pi^{+}$, $\pi^{0}$ or $\pi^{-}$;][]{article:Hagiwara_etal:2002} and the other resonance,
res(1600), is assumed to decay to a nucleon and two pions. The extension of model A which includes
the baryon resonances is from here on referred to as adjusted model A.

Other necessary adjustments to model A forced by the introduction of the baryon resonance
contributions is described in \citet[section 3]{article:Kamae_etal:2006}. This includes the
adjustment of the non-diffractive inelastic $p$-$p$ cross section to accommodate for the resonances
while not exceeding the total inelastic cross section. The total inelastic $p$-$p$ cross section is
shown in Figure \ref{fig:pp_cross_sections} together with the four component cross sections of
adjusted model A. 

Due to paucity of experimental data, $\alpha$-$p$, $p$-He and $\alpha$-He have not been included in
this work. Near the Earth about 7\% of the CR flux is $\alpha$-particles
\citep{book:Schlickeiser:2002} and the ISM contains about 10\% He by number. Both the
$\alpha$-particle and the He nucleus can be approximated as four individual nucleons. The error from
such an approximation is expected to be less than 10\% for high-energy gamma rays. Fermi motion of
nucleons and multiple nucleonic interactions in the nucleus affect the pion production near the
threshold and in the resonance region \citep[$T_{p}<3$ GeV;][]{article:Crawford_etal:1980,
article:Martensson_etal:2000}. This will enhance the pion multiplicity below 100 MeV. The need for
separate treatment of interactions such as $p$-He, $\alpha$-$p$, and $\alpha$-He is acknowledged.

\section{Monte Carlo Simulations}
The parameterization of the angular distribution of gamma rays is derived from Monte Carlo
simulations on the adjusted proton-proton interaction model A as given in
\citet{article:Kamae_etal:2006}. Events were generated for each of the four components, the
non-diffractive interaction, the diffraction dissociation process and the two resonance excitation
processes in the following way. For the non-diffractive interaction in the high energy range
($T_{p}>52.6$ GeV) Pythia 6.2 \citep{article:Sjostrand_etal:2001} was used with the option for
multiparton-level scaling violation
\citep{article:SjostrandSkands:2004}.\footnote{See http://cepa.fnal.gov/CPD/MCTuning1 and
http://www.phys.ufl.edu/$\sim$rfield/cdf} This was complemented with the parameterization of
inclusive pion cross sections by \citet{article:Blattnig_etal:2000} in the low energy range
($T_{p}<52.6$ GeV). The diffraction dissociation process was simulated with a Monte Carlo code by T.
Kamae (2004, personal communications)\footnote{It is acknowledged that the latest version of Pythia
includes the diffractive interaction and that the code used here agrees with Pythia. The Monte Carlo
code is available upon request.} and the resonance excitation components were simulated with Monte
Carlo codes by T. Kamae (2005, personal communications). 

For each of the four components mentioned above, events were generated for discrete proton kinetic
energies (0.488 GeV $\leq T_{p} \leq$ 512 TeV) taken from a geometrical series
\begin{equation}
T_{p}=1000\cdot 2^{(i-22)/2}\ \mathrm{GeV},\ i=0,\ldots,40.
\end{equation}
Each proton kinetic energy, $T_{p}$, represents a bin covering $2^{-0.25}T_{p}$ to $2^{0.25}T_{p}$.
The addition of the resonances to the model required an increased sampling frequency near the pion
production threshold and events were also generated for $T_{p}=0.58$ GeV and 0.82 GeV. Events were
not generated for proton energies where the component cross section is very small or zero.

\subsection{Pion Transverse Momentum}
The above described Monte Carlo simulations have been verified to agree with experimental data
available for pions. The inclusive $\pi^{0}$ cross section was verified to agree with the
experimental one and the simulations were ensured to reproduce the distributions of pion kinetic
energy in the $p$-$p$ center-of-mass (CM) system in the resonance region
\citep{article:Kamae_etal:2006}.

\begin{figure}
\begin{center}
\scalebox{1.0}{\plotone{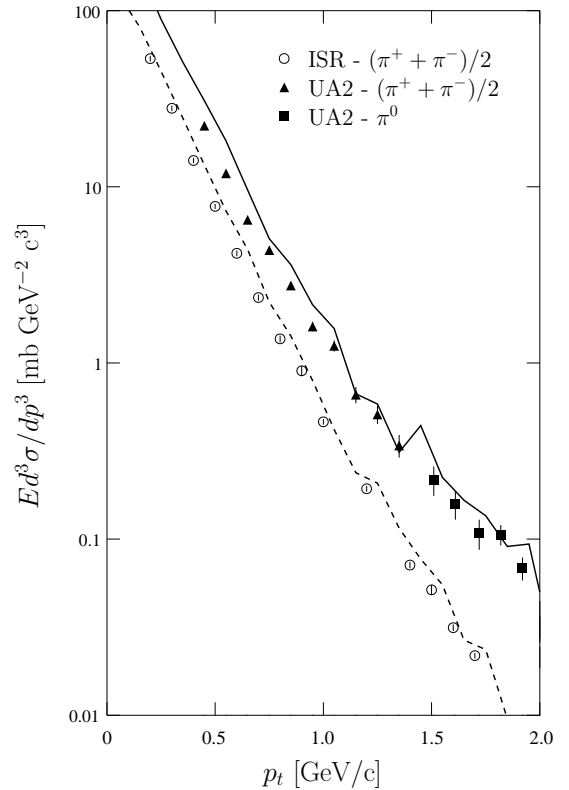}}
\end{center}
\caption{Experimental invariant cross section $Ed^{3}\sigma/dp^{3}$ at production angle
$\theta_{\mathrm{cms}}=90^{\circ}$ and beam energy $\sqrt{s}=540$ GeV for $(\pi^{+}+\pi^{-})/2$
(filled triangles) and $\pi^{0}$ (filled squares) measured by the UA2 collaboration
\citep[$\bar{p}$-$p$ collider experiment;][]{article:Banner_etal:1982, article:Banner_etal:1983}
and beam energy $\sqrt{s}=53$ GeV for $(\pi^{+}+\pi^{-})/2$ (open circles) measured at the ISR
\citep[$p$-$p$ collider; ][]{article:Alper_etal:1975b} together
with the $\pi^{0}$ invariant cross section calculated from Monte Carlo simulations in this work at
$T_{p}=181$ TeV ($\sqrt{s}=582$ GeV, solid line) and $T_{p}=1.41$ TeV ($\sqrt{s}=51.5$ GeV, dashed
line).}
\label{fig:pi_pT_dists}
\end{figure}

For this work, the $\pi^{0}$ transverse momentum distribution was compared with those measured with
accelerator experiments. Figure \ref{fig:pi_pT_dists} shows the invariant $\pi^{0}$ cross section,
$Ed^{3}\sigma/dp^{3}$, at production angle $\theta_{\mathrm{cms}}=90^{\circ}$ for proton kinetic
energies $T_{p}=1.41$ TeV ($\sqrt{s}=51.5$ GeV) and 181 TeV ($\sqrt{s}=582$ GeV) together with
experimental data for $\pi^{\pm}$ measured at the ISR at $\sqrt{s}=53$ GeV
\citep{article:Alper_etal:1975b} and by the UA2 collaboration at $\sqrt{s}=540$ GeV
\citep{article:Banner_etal:1982, article:Banner_etal:1983}. One must note that the ISR was a $p$-$p$
collider and that UA2 was a $\bar{p}$-$p$ collider experiment. The distributions follow the
expected exponential form for small $p_{t}$.

\begin{figure}
\begin{center}
\scalebox{1.0}{\plotone{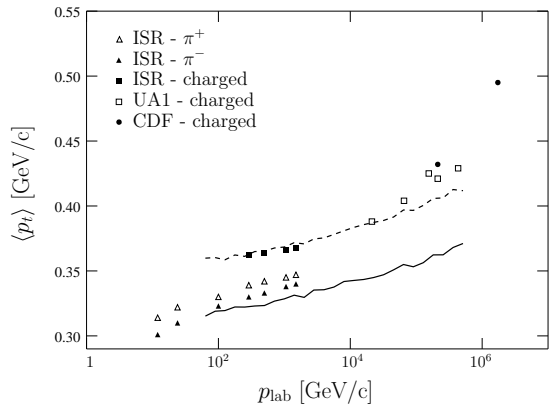}}
\end{center}
\caption{Average transverse momentum, $\expval{p_{t}}$, for production of pions and all charged
particles ($\pi^{\pm}$, $K^{\pm}$ and protons) versus the laboratory momentum. Data points are for
$\pi^{+}$ (open triangles), $\pi^{-}$ (filled triangles) and all charged particles (filled squares)
from ISR, all charged particles (filled diamonds) from UA1 and all charged particles (filled
circles) from CDF. ISR data are from $p$-$p$ collider experiments and are taken from
\citet{article:Rossi_etal:1975} and UA1 and CDF data are from $\bar{p}$-$p$ collider experiments and
are taken from \citet{article:Abe_etal:1988}. Lines are for $\pi^{0}$ (solid line) and all charged
particles (dashed line) from Monte Carlo simulations in this work.}
\label{fig:average_pT}
\end{figure}

In addition, the energy dependence of the average transverse momentum, $\expval{p_{t}}$, calculated
from Monte Carlo event data was compared with that from ISR experiments. In accelerator experiments
it is difficult to measure $\pi^{0}$ directly but one can expect that
$\expval{p_{t}[\pi^{0}]}\simeq\expval{p_{t}[\pi^{\pm}]}$ \citep{article:Alner_etal:1987}.
$\expval{p_{t}[\pi^{0}]}$ was calculated without any fitting and Figure \ref{fig:average_pT} shows
the average transverse momentum as a function of the proton momentum in the laboratory frame.
\citet{article:Rossi_etal:1975} estimated the error on $\expval{p_{t}[\pi^{\pm}]}$ to be about 
$10\%$ and $\expval{p_{t}[\pi^{0}]}$ calculated here is within this error margin.

At very high proton momentum experimental data is in general limited to the average transverse
momentum of charged particles, $\expval{p_{t}[\mathrm{charged}]}$, where charged particles include
charged pions and kaons and protons. Again, without fitting, $\expval{p_{t}[\mathrm{charged}]}$ was
calculated from Monte Carlo event data and compared with experimental data measured at the ISR and
by the UA1 and CDF collaborations.\footnote{Note again that ISR was a $p$-$p$ collider and that UA1
and CDF are $\bar{p}$-$p$ collider experiments.} As can been seen in Figure \ref{fig:average_pT} the
difference is on the order of a few percent except at very high energies. The discrepancy in average
transverse momentum at high proton momentum is inherent to Pythia 6.2 which has not been fine-tuned
to the highest energy data available from CDF (T. Sj\"ostrand 2007, personal communications).

\subsection{Angular Distribution of Gamma Rays}
The Monte Carlo simulations generate data on momentum, $\mathbf{p}=(p_{x},p_{y},p_{z})$, and total
energy, $E$, for each gamma-ray photon. Since the incident proton direction in the simulations is
along the z-axis the transverse momentum, $p_{t}$, is simply
\begin{equation}\label{eq:pT}
p_{t}=\sqrt{p_{x}^{2}+p_{y}^{2}}
\end{equation}
and this was calculated for each simulated event. Events were then binned in 2D histograms,
\begin{equation}
\frac{\Delta^{2}N}{\Delta\log(E)\Delta p_{t}},
\end{equation}
over total energy, $E$, and transverse momentum, $p_{t}$, with one histogram per proton kinetic
energy. Bins widths were taken to be $\Delta\log(E)=0.05$ and $\Delta p_{t}=10$ MeV/c.

Normalization to the proton-proton inelastic cross section, $\sigma_{pp}$
\citep[given by eqs. (1) through (4) in][]{article:Kamae_etal:2006}, and per proton-proton
interaction gives the differential cross section
\begin{equation}
\frac{\Delta^{2}\sigma}{\Delta\log(E)\Delta p_{t}}=\frac{\sigma_{pp}}{N_{pp}}\frac{\Delta^{2}N}{\Delta\log(E)\Delta p_{t}},
\end{equation}
where $N_{pp}$ is the number of proton-proton events simulated and $\Delta N$ is the number of gamma
rays in a given bin. This differential cross section is a representation of the angular distribution
of gamma rays.

\section{Parameterization of Gamma-Ray Transverse Momentum Distributions}
For each proton kinetic energy, $T_{p}$, the transverse momentum distribution is parameterized as
\begin{equation}\label{eq:pT_rep}
\frac{\Delta^{2}\sigma}{\Delta\log(E)\Delta p_{t}}=p_{t}F(p_{t}, x)F_{\rm{kl}}(p_{t},x),
\end{equation}
where $x=\log{(E\mathrm{[GeV]})}$, $F(p_{t},x)$ is the function representing the differential cross
section $\Delta\sigma/\Delta p_{t}^{2}$ and $F_{\rm{kl}}(p_{t},x)$ is used to approximately enforce
the energy-momentum conservation. Assuming axial symmetry around the $p_{\parallel}$ axis, phase
space is proportional to $dp_{t}^{2}dp_{\parallel}=2p_{t}dp_{t}dp_{\parallel}$, which gives the
extra factor of $p_{t}$ in equation (\ref{eq:pT_rep}).

The function $F_{\rm{kl}}(p_{t},x)$ enforcing the energy-momentum conservation is taken to be
\begin{equation}
F_{\rm{kl}}(p_{t},x)=\frac{1}{\exp{(W(p_{t} - L_{p}))} + 1},
\end{equation}
where $W=75$ and
\begin{equation}
L_{p}=\left\{\begin{array}{ll}
0.0976 + 0.670\exp(1.81x) & x < -1 \\
- 0.793 + \exp(0.271(x + 1) \\
\ \ \ \ + 0.363(x + 1)^{2}) & -1 \leq x < 0.5 \\
2.5 & x \geq 0.5,
\end{array}\right.
\end{equation}
with $x=\log{(E\mathrm{[GeV]})}$.

In contrast to the parameterization of inclusive cross sections by \citet{article:Kamae_etal:2006},
where the non-diffractive and the diffraction contribution were treated separately, the two are here
merged to one contributing component. This is well justified in astrophysical contexts. The new
component is from here on referred to as the non-resonance component.

The $p_{t}$ distribution is given by 2D histograms, one histogram per proton kinetic energy,
$T_{p}$, and component: non-resonance, $\Delta(1232)$, and ${\rm{res}}(1600)$. Each histogram is
fitted in slices along $p_{t}$, i.e. each slice $\Delta\sigma/\Delta p_{t}^{2}$ covers one bin of
$\Delta \log(E)$. Note that $\Delta\sigma/\Delta p_{t}^{2}$ does not imply integrating over
$\log(E)$.

\begin{deluxetable*}{llr}
\tablecolumns{3}
\tablewidth{0pc}
\tablecaption{Parameters describing transverse momentum distributions}
\tablehead{
\colhead{Parameters} &
\multicolumn{2}{c}{Formulae as functions of the proton kinetic energy, $y=\log{(T_{p}\mathrm{[TeV]})}$}\\
\colrule
\multicolumn{3}{c}{Non-resonance, eq. (\ref{eqn:a1})}}
\startdata
$a_{10}$\dotfill & $0.043775 + 0.010271\exp(-0.55808y)$ \\
$a_{11}$\dotfill & $0.8$ \\
$a_{12}$\dotfill & $0.34223 + 0.027134y - 0.0089229y^{2} + 4.9996\times 10^{-4}y^{3}$ \\
$a_{13}$\dotfill & $-0.20480 + 0.013372y + 0.13087\exp{(0.0044021(y - 11.467)^{2})}$ \\
$a_{14}$\dotfill & $a_{1}(x = -0.75)$ \\
\cutinhead{$\Delta(1232)$, eq. (\ref{eqn:bi})}
$b_{10}$\dotfill & $18.712 + 18.030y + 5.8239y^{2} + 0.62728y^{3}$ \\
$b_{11}$\dotfill & $612.61 + 404.80y + 67.406y^{2}$ \\
$b_{12}$\dotfill & $98.639 + 96.741y + 31.597y^{2} + 3.4567y^{3}$ \\
$b_{13}$\dotfill & $-208.38 - 183.65y - 53.283y^{2} - 5.0470y^{3}$ \\
\\
$b_{20}$\dotfill & $0.21977 + 0.064073x$ \\
$b_{21}$\dotfill & $3.3187\times10^{3} + 3463.4y + 1.1982\times10^{3}y^{2} + 136.71y^{3}$ \\
$b_{22}$\dotfill & $91.410 + 91.613y + 30.621y^{2} + 3.4296y^{3}$ \\
$b_{23}$\dotfill & $-521.40 - 529.06y - 178.49y^{2} - 19.975y^{3}$ \\
\cutinhead{res(1600), eq. (\ref{eqn:ci})}
$c_{10}$\dotfill & $-1.5013 -1.1281y - 0.19813y^{2}$ \\
$c_{11}$\dotfill & $-33.179 - 22.496y - 3.3108y^{2}$ \\
$c_{12}$\dotfill & $116.44 + 122.11y + 42.594y^{2} + 4.9609y^{3}$ \\
$c_{13}$\dotfill & $-545.77 - 574.80y - 201.25y^{2} - 23.400y^{3}$ \\
\\
$c_{20}$\dotfill & $0.68849 + 0.36438y + 0.047958y^{2}$ \\
$c_{21}$\dotfill & $-1.6871\times 10^{4} - 1.7412\times 10^{4}y - 5.9648\times 10^{3}y^{2} - 679.27y^{3}$ \\
$c_{22}$\dotfill & $-88.565 - 94.034y - 33.014y^{2} - 3.8205y^{3}$ \\
$c_{23}$\dotfill & $1.5141\times 10^{3} + 1.5757\times 10^{3}y + 544.20y^{2} + 62.446y^{3}$ \\
\enddata
\label{table:pT_params}
\end{deluxetable*}

For the non-resonance component $\Delta\sigma/\Delta p_{t}^{2}$ is expected to follow an exponential
form 
\begin{equation}\label{eqn:Fnr}
F_{\rm{nr}}(p_{t},x)=a_{0}\exp{\left(-\frac{p_{t}}{a_{1}}\right)}.
\end{equation}
Parameter $a_{1}$ gives the shape of the differential cross section and $a_{0}$ gives the absolute
normalization. When integrating over $p_{t}$ one should recover the inclusive cross section
$\Delta\sigma/\Delta\log(E)$, i.e.
\begin{equation}\label{eq:pt_norm}
\int_{0}^{\infty}\frac{\Delta\sigma}{\Delta p_{t}^{2}}dp_{t}=\frac{\Delta\sigma}{\Delta\log(E)}.
\end{equation}
Thus, $a_{0}$ is taken such that
\begin{equation}
a_{0}\int_{0}^{\infty}p_{t}\exp{\left(-\frac{p_{t}}{a_{1}}\right)}dp_{t}=\frac{\Delta\sigma}{\Delta\log(E)}
\end{equation}
which gives
\begin{equation}
a_{0}=\frac{1}{a_{1}^{2}}\frac{\Delta\sigma}{\Delta\log(E)}
\end{equation}
and $\Delta\sigma/\Delta\log(E)$ is calculated using the parameterization of the inclusive cross
section by \citet{article:Kamae_etal:2006}.

Parameter $a_{1}$ is a function of both the gamma-ray energy, $E$, and the proton kinetic energy,
$T_{p}$. It is first fitted as a function of $x=\log{(E\mathrm{[GeV]})}$ for each simulated proton
kinetic energy. The formula describing $a_{1}$ is
\begin{equation}\label{eqn:a1}
a_{1}(x) = \left\{\begin{array}{ll}
a_{10}\exp{(-a_{11}(x + a_{12})^{2})} & x \leq -0.75,\\
a_{13}(x + 0.75) + a_{14} & x > -0.75.\\
\end{array}\right.
\end{equation}
The parameters $a_{1i}$ ($i=0,\ldots,4$) are then given by functions of the proton kinetic energy,
which are listed in Table \ref{table:pT_params}.

For the baryon resonance components $\Delta\sigma/\Delta p_{t}^{2}$ will not follow the exponential
form. Instead, $F(p_{t},x)$ is fitted to a Gaussian form
\begin{equation}\label{eqn:Fdelta}
F_{\Delta(1232)}(p_{t},x)=b_{0}\exp{\left(-\frac{(p_{t}-b_{1})^{2}}{b_{2}}\right)}
\end{equation}
and
\begin{equation}\label{eqn:Fres}
F_{\rm{res(1600)}}(p_{t},x)=c_{0}\exp{\left(-\frac{(p_{t}-c_{1})^{2}}{c_{2}}\right)}.
\end{equation}
With the requirement that the integral over $p_{t}$ should recover the inclusive cross section (eq.
\ref{eq:pt_norm})
\begin{eqnarray}
b_{0}&=&2(b_{1}\sqrt{\pi b_{2}}(\mathrm{erf}(b_{1}/\sqrt{b_{2}})+1)+\nonumber\\
&+&b_{2}\exp(-b_{1}^{2}/b_{2}))^{-1}\frac{\Delta\sigma}{\Delta\log(E)}
\end{eqnarray}
and
\begin{eqnarray}
c_{0}&=&2(c_{1}\sqrt{\pi c_{2}}(\mathrm{erf}(c_{1}/\sqrt{c_{2}})+1)+\nonumber\\
&+&c_{2}\exp(-c_{1}^{2}/c_{2}))^{-1}\frac{\Delta\sigma}{\Delta\log(E)}.
\end{eqnarray}
Again, the parameters $b_{i}$ and $c_{i}$ ($i=1,2$) are functions of both $E$ and $T_{p}$ and the
same procedure is followed for them, with
\begin{equation}\label{eqn:bi}
b_{i}(x) = b_{i0}\exp{\left(-b_{i1}\left(\frac{x - b_{i2}}{1.0 + b_{i3}(x - b_{i2})}\right)^{2}\right)}
\end{equation}
and
\begin{equation}\label{eqn:ci}
c_{i}(x) = c_{i0}\exp{\left(-c_{i1}\left(\frac{x - c_{i2}}{1.0 + c_{i3}(x - c_{i2})}\right)^{2}\right)},
\end{equation}
for $x < 0.5$ and $b_{i}(x)=0$ for $x \geq x_{b}$ and $c_{i}(x)=0$ for $x \geq x_{c}$, with
\begin{eqnarray}
x_{b} & = & 0.81(y + 3.32) - 0.5\\
x_{c} & = & 0.82(y + 3.17) - 0.25,
\end{eqnarray} 
where $y=\log(T_{p}[\mathrm{TeV}])$. These limits of $b_{i}$ and $c_{i}$ were introduced to control
artifacts near the kinematical limits. The parameters $b_{ij}$ and $c_{ij}$ ($j=0,\ldots,4$) are
listed in Table \ref{table:pT_params} as functions of the proton kinetic energy.

\begin{figure*}
\begin{center}
\scalebox{0.85}{\plottwo{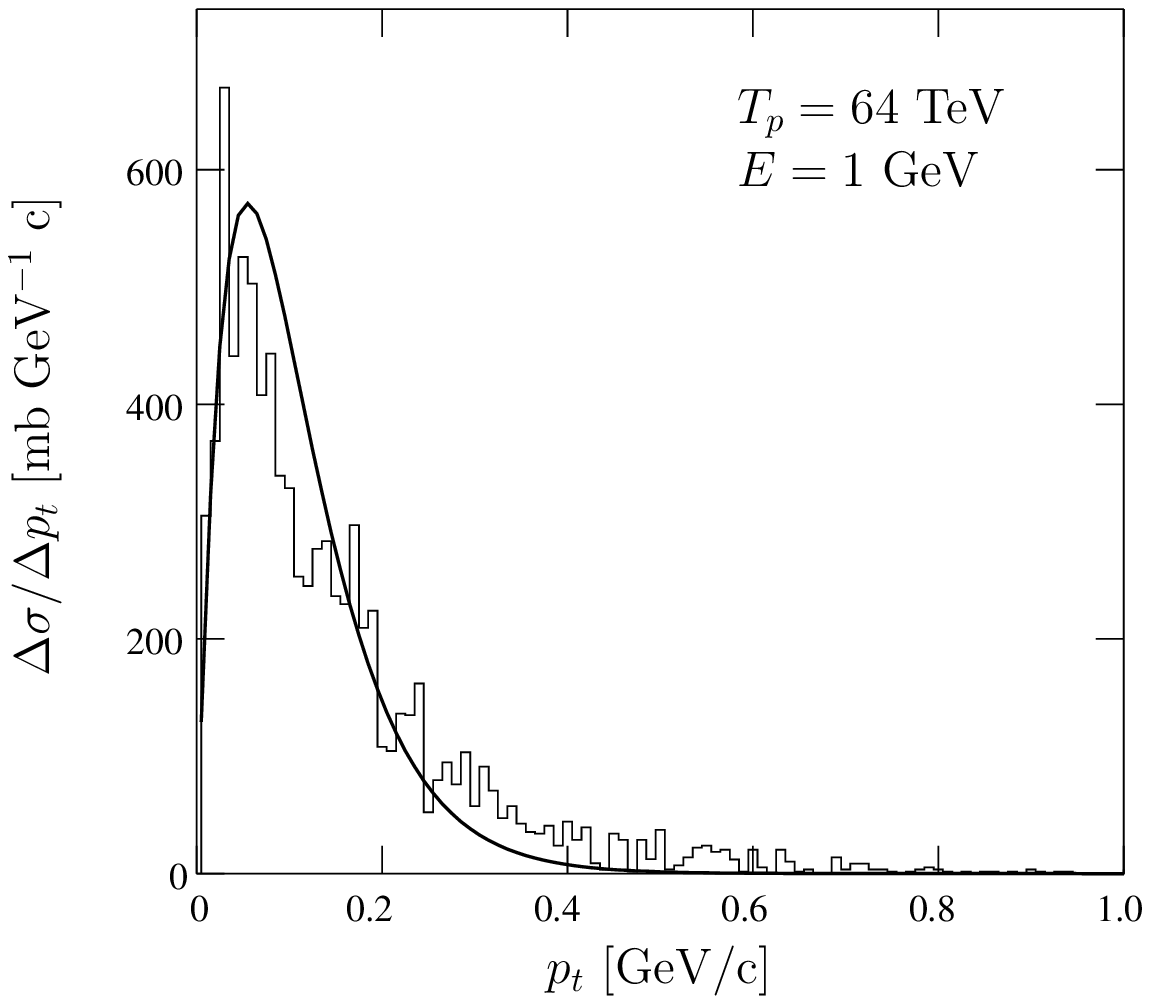}{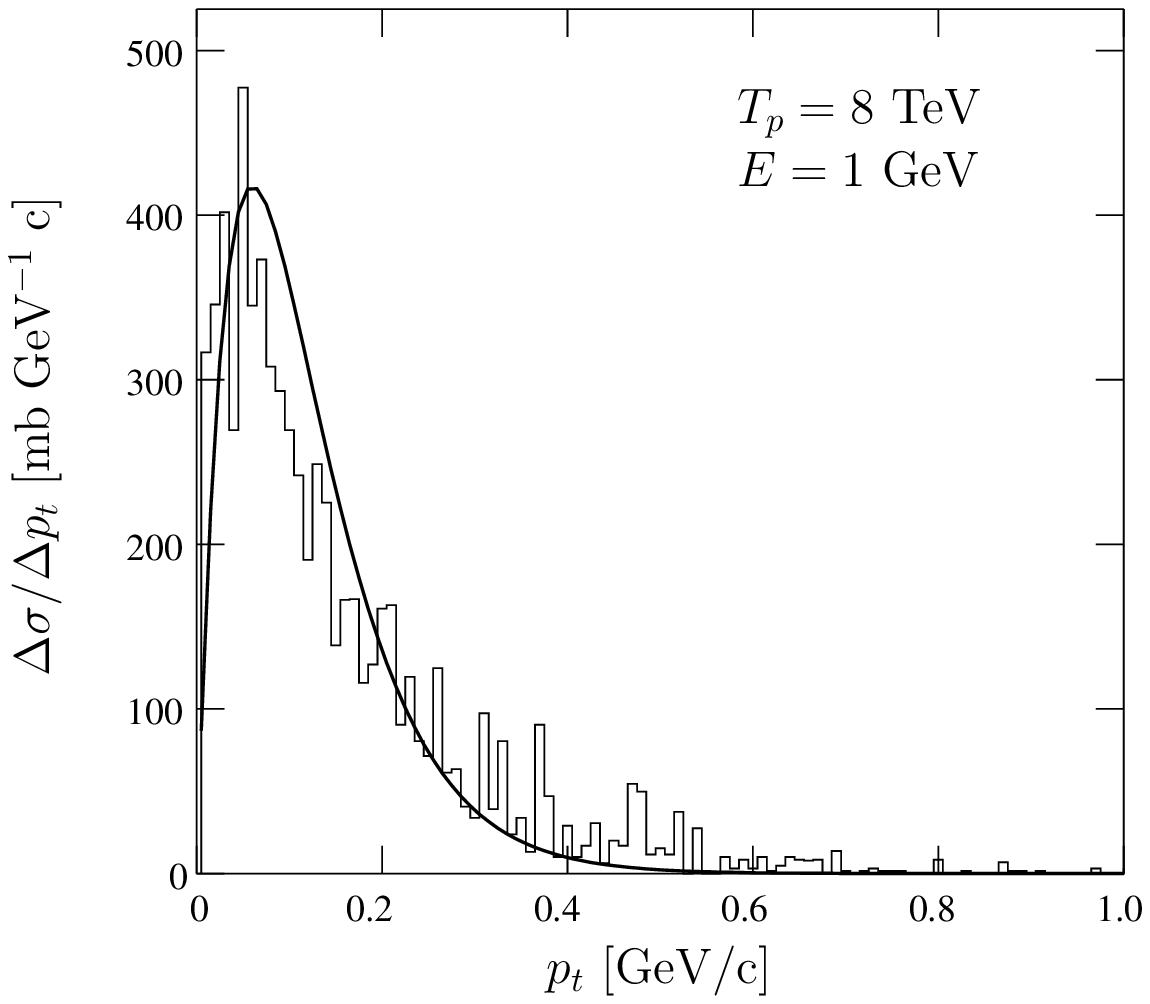}}
\scalebox{0.85}{\plottwo{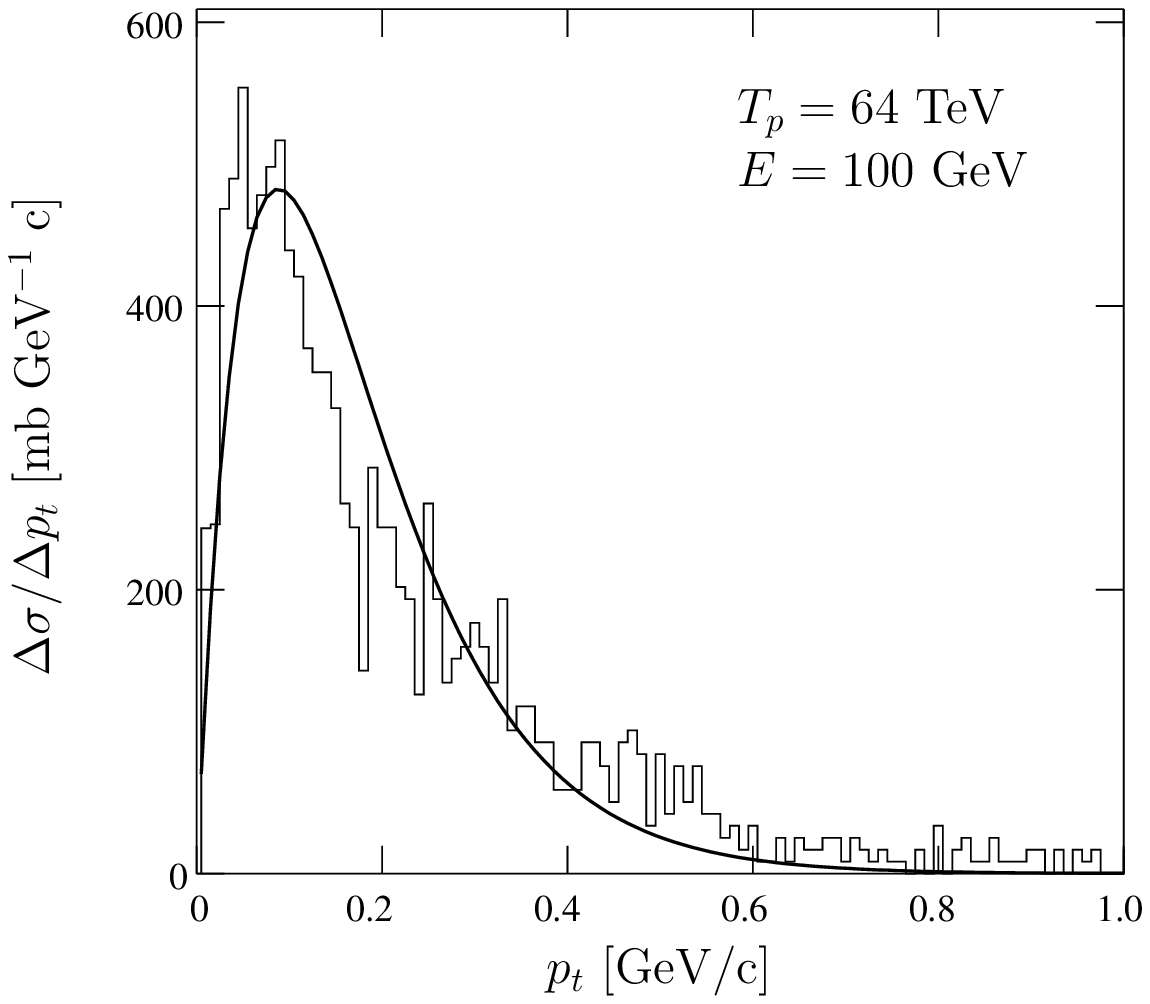}{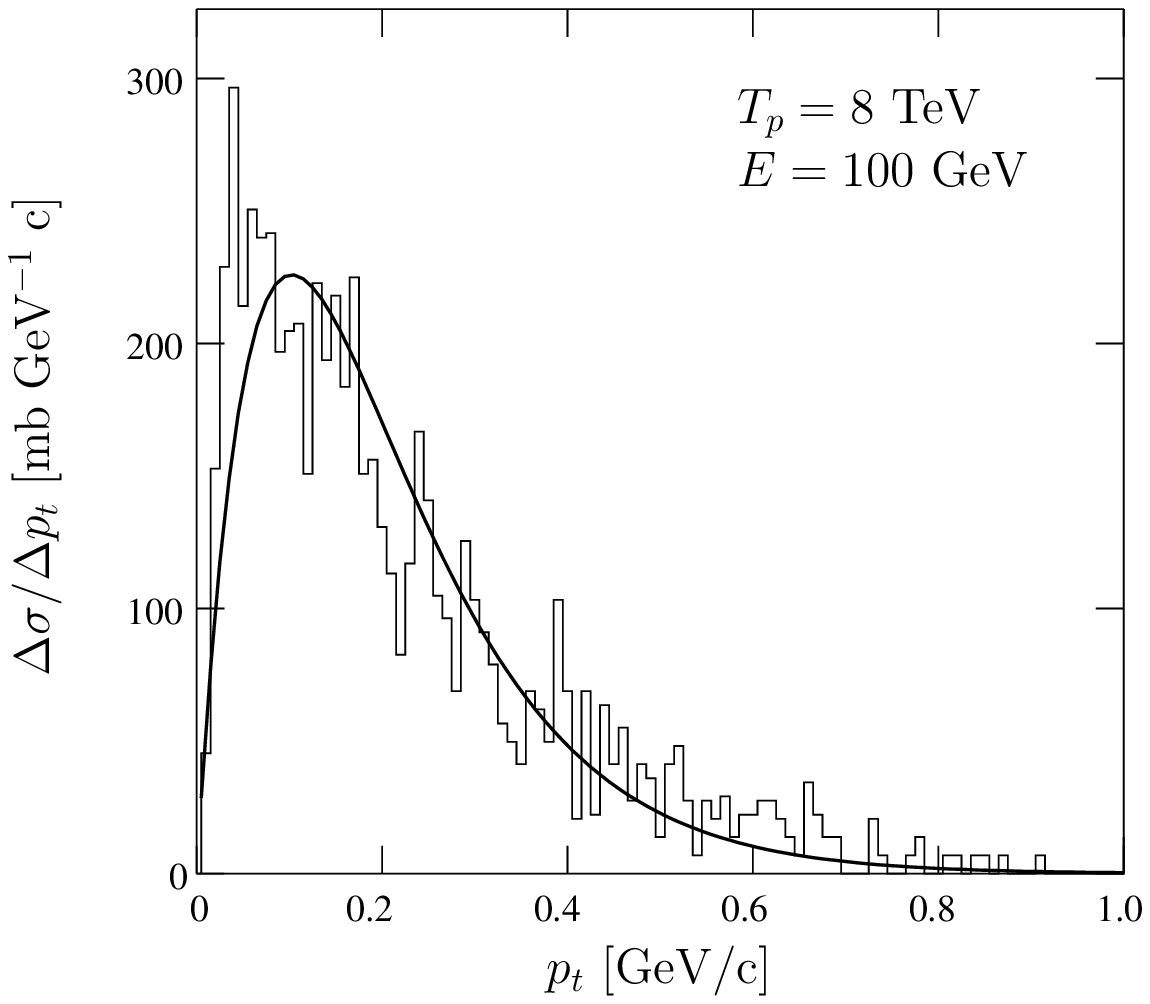}}
\scalebox{0.85}{\plottwo{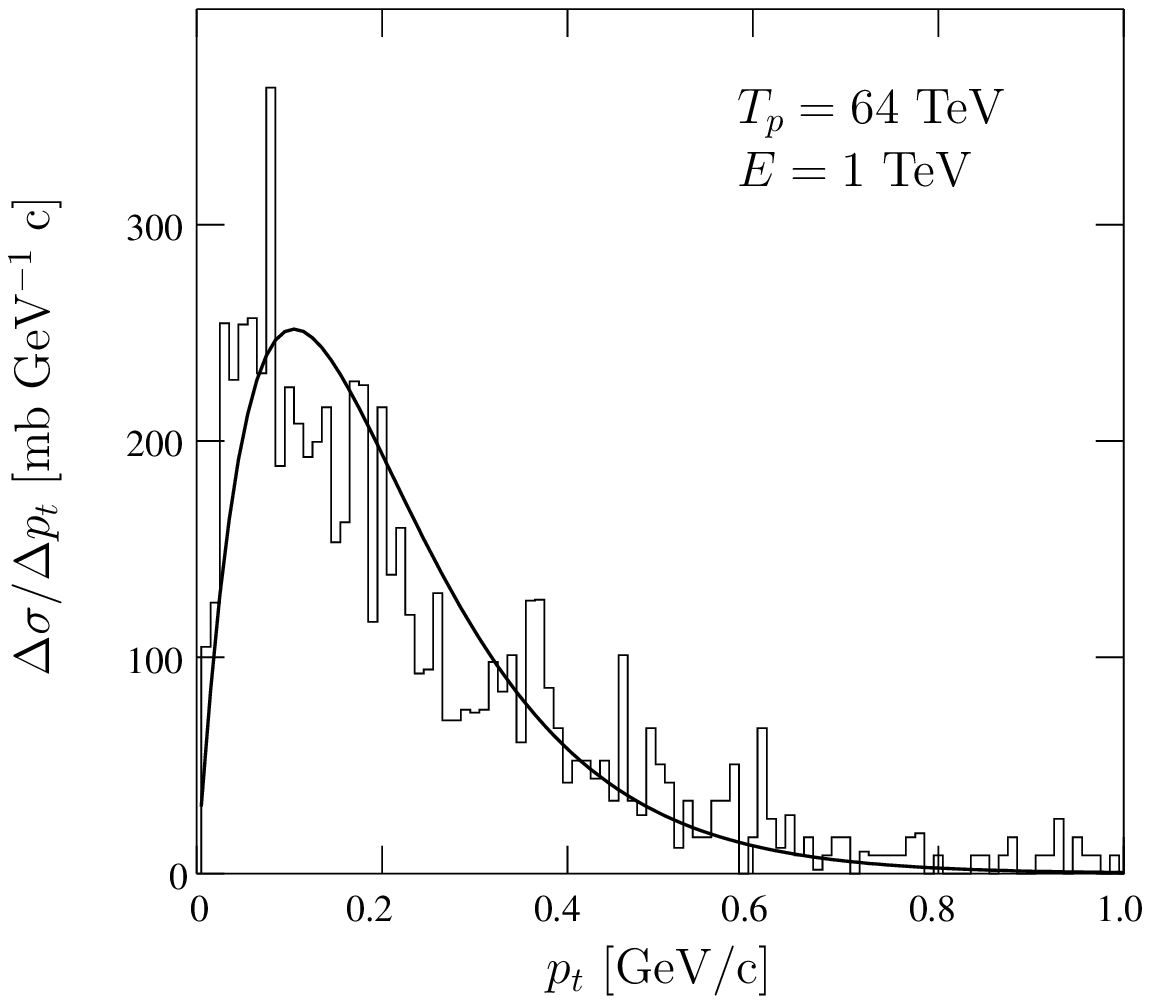}{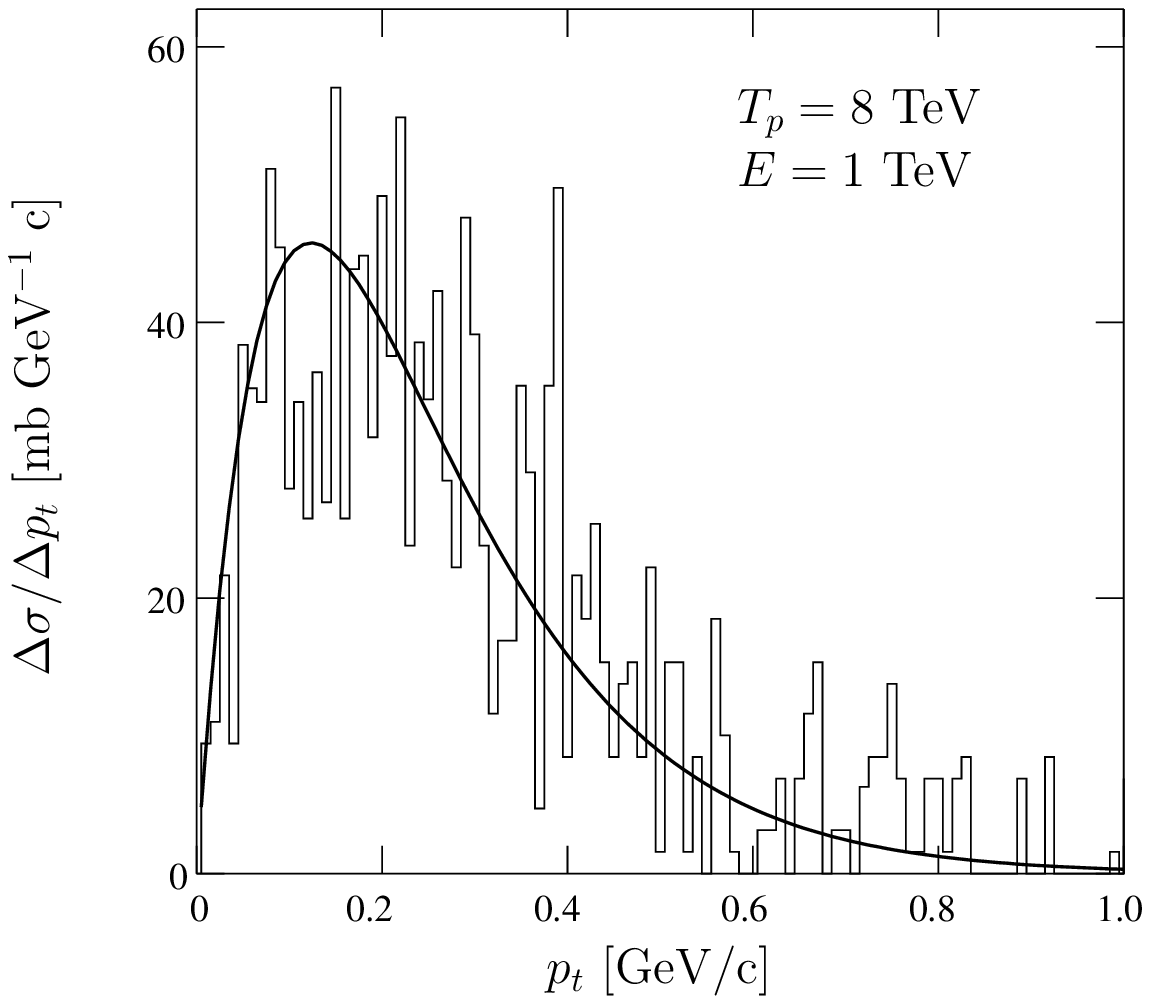}}
\end{center}
\caption{Gamma-ray differential cross section $\Delta\sigma/\Delta p_{t}$ for the non-resonance
contribution calculated using the parameterization (thick solid line) and superimposed with the
Monte Carlo simulated cross section (thin histogram). Panels on the left-hand side are for proton
kinetic energy $T_{p}=64$ TeV and panels on the right-hand side are for $T_{p}=8$ TeV. Rows are for
gamma-ray energy $E=1$ GeV (top), 100 GeV (middle), and 1 TeV (bottom).}
\label{fig:pT_dists_nr}
\end{figure*}

\begin{figure}
\begin{center}
\scalebox{0.85}{\plotone{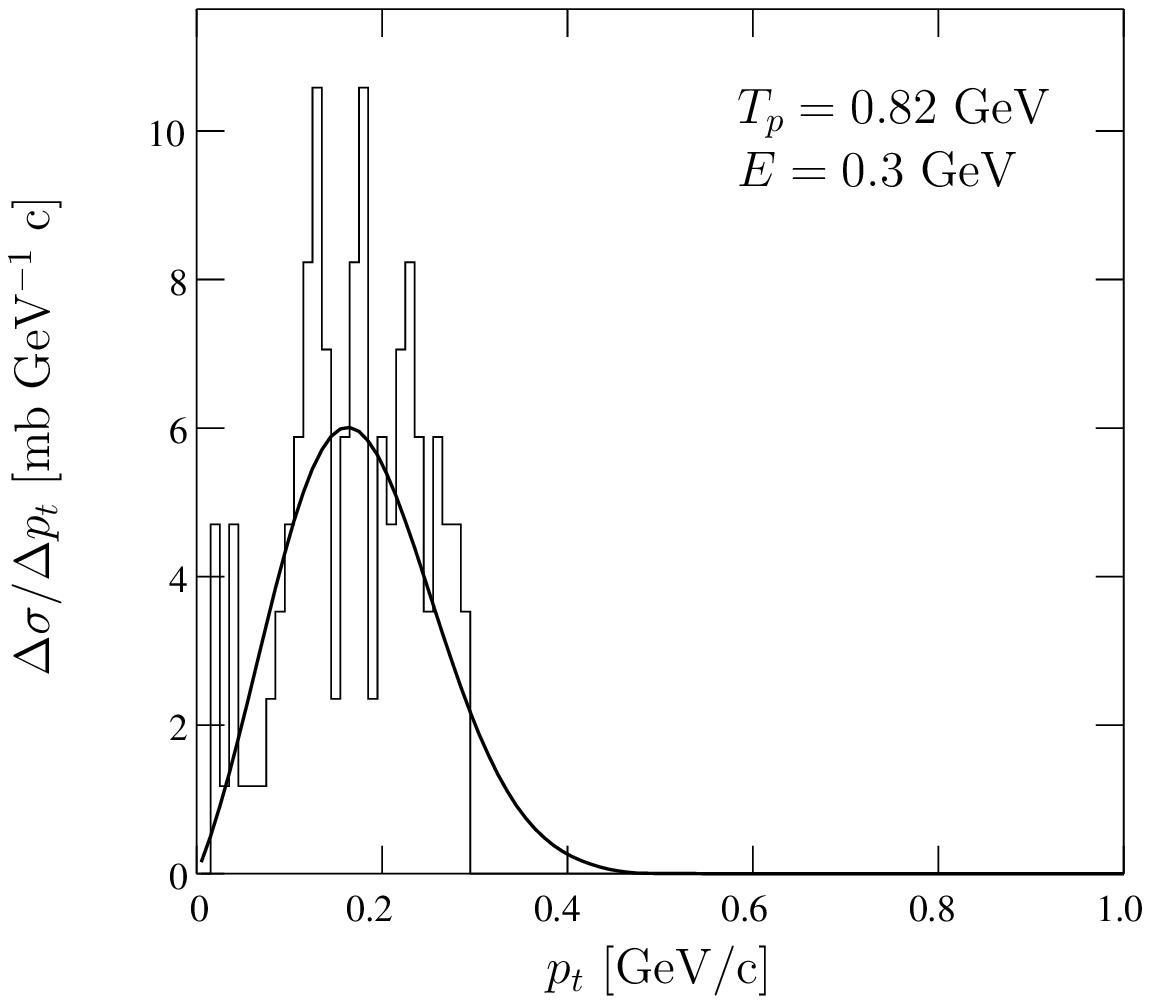}}
\scalebox{0.85}{\plotone{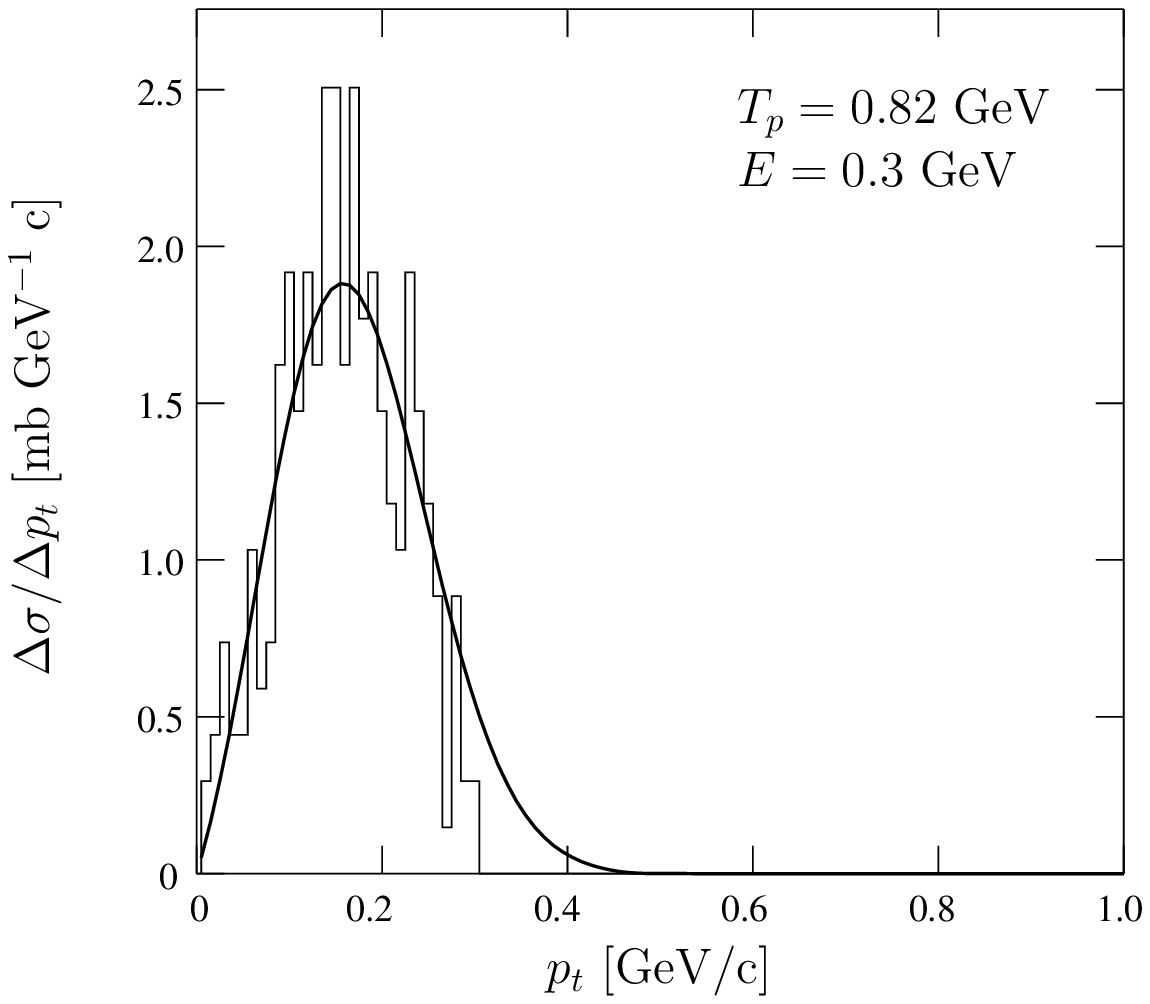}}
\end{center}
\caption{Gamma-ray differential cross section $\Delta\sigma/\Delta p_{t}$ for the two resonance
contributions, $\Delta(1232)$ and res(1600), calculated using the parameterization (thick solid
line) and superimposed with the Monte Carlo simulated cross section (thin histogram) for proton
kinetic energy $T_{p}=0.82$ GeV and $E=0.3$ GeV. The top panel is for for the $\Delta(1232)$
resonance and the bottom panel is for the res(1600) resonance.}
\label{fig:pT_dists_res}
\end{figure}

Figure \ref{fig:pT_dists_nr} shows the gamma-ray differential cross section
$\Delta\sigma/\Delta p_{t}$ for the non-resonance contribution calculated using the above described
parameterization for proton kinetic energies $T_{p}=64$ TeV and 8 TeV and gamma-ray energies $E=1$
GeV, 100 GeV and 1 TeV. The plots show how the average transverse momentum, $\expval{p_{t}}$,
increases with increasing gamma-ray energy. Figure \ref{fig:pT_dists_res} shows
$\Delta\sigma/\Delta p_{t}$ for the two resonance contributions, $\Delta(1232)$ and res(1600),
calculated at proton kinetic energy $T_{p}=0.82$ GeV and gamma-ray energy $E=0.3$ GeV. Superimposed
in both figures are the differential cross sections from the Monte Carlo simulations. The agreement
is in general good except near the higher and lower kinematical limits where low statistics in the
Monte Carlo simulations limits accuracy of the fit.

\section{Application of Formulae}
The parameterized model of stable secondary particle spectra by \citet{article:Kamae_etal:2006} was
used to predict differences in the diffuse gamma-ray spectrum from the Galactic ridge compared with
the scaling models implemented in GALPROP. The present model finds its application in scenarios
where the gamma-ray spectrum is expected to be angular dependent, such as AGN jets and GRBs, but
SNRs may also fall into this category. The highest-energy CR escape the forward shock almost
unidirectionally giving rise to beaming in SNRs.

To demonstrate the use of the parameterized formulae for gamma-ray $p_{t}$ distributions, the
gamma-ray spectrum has been calculated for two different cases; the first is a pencil beam of
protons following a power law of index 2.0 and the other is a fanned proton jet with a Gaussian
angular profile impinging on the surrounding matter. 

\subsection{Pencil Beam of Protons}
\begin{figure}
\begin{center}
\scalebox{1.0}{\plotone{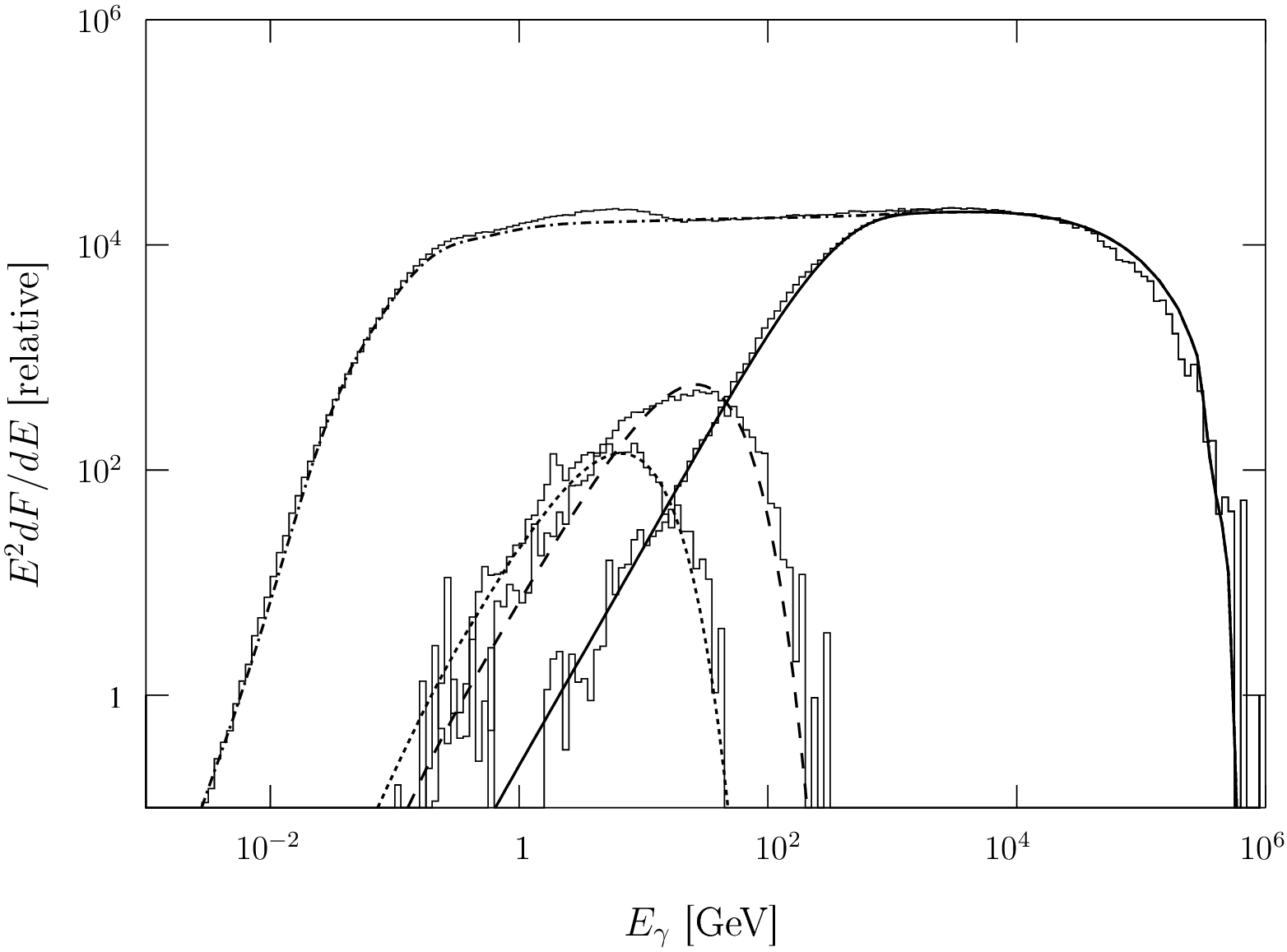}}
\end{center}
\caption{Gamma-ray spectra from a pencil beam of protons observed from three different angles,
$\theta=0^{\circ}$ (solid), $0^{\circ}\!\!.5$ (dashed), and $2^{\circ}$ (dotted) calculated using
the parametric model. The spectra are integrated over the annular portion $(\theta, \theta+d\theta)$
of width $d\theta=2'$. Included is also the spectrum integrated over the entire phase space
(dash-dotted). Histograms are the corresponding Monte Carlo spectra. The protons in the beam are
assumed to follow a power law in kinetic energy, $T_{p}$, with index 2.0 and extending up to
$T_{p}=512$ TeV. Fluctuations in the histograms are due to low event statistics}
\label{fig:pencil_beam_spectra}
\end{figure}

Consider a beam of protons along the z-axis with no spatial extension in the x-y plane, i.e. a
pencil beam. The energy distribution of protons is assumed to be a power law, $dN/dE=T_{p}^{-s}$,
with index $s=2.0$ and extending up to $T_{p}=512$ TeV. The gamma-ray spectrum, $E^{2}dF/dE$, is
calculated for three different observation angles $\theta=0^{\circ}$ (head on), $0^{\circ}\!\!.5,$
and $2^{\circ}$ relative to the beam axis. The spectra, which are shown in Figure
\ref{fig:pencil_beam_spectra}, are integrated over the annular portion $(\theta, \theta+d\theta)$ of
width $d\theta=2'$. The absolute normalization is relative to the density and distribution of target
protons. For comparison the spectrum integrated over the entire phase space is also plotted in the
figure. As can be seen in the figure, the gamma-ray emission is peaked in the very forward
direction. When the viewing angle is increased, the peak of the spectrum is shifted to lower
gamma-ray energies and the the flux decreases rapidly. Fluctuations in the histograms are due to low
event statistics.

\subsection{Fanned Proton Jet}
\begin{figure}
\begin{center}
\scalebox{1.0}{\plotone{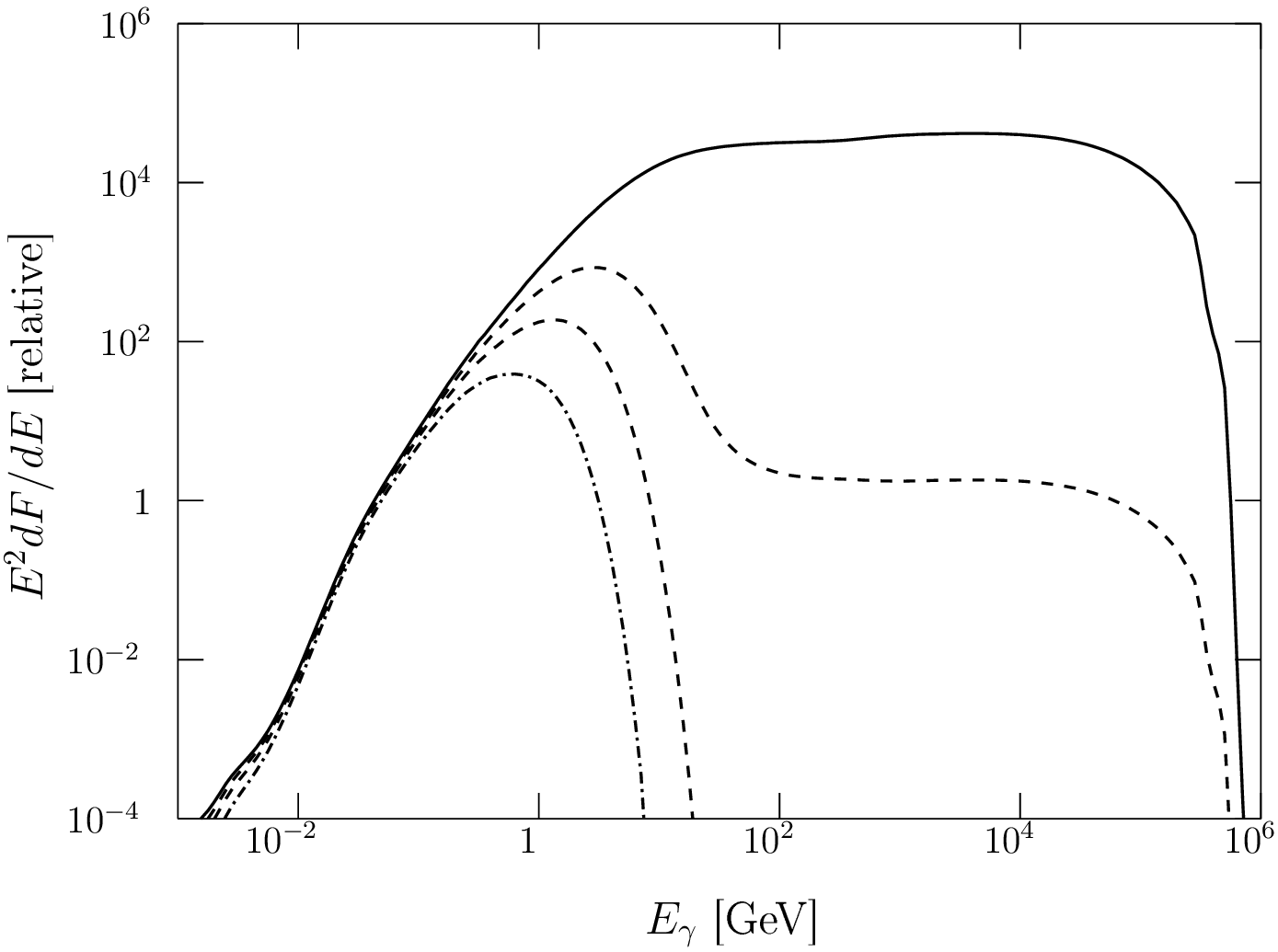}}
\end{center}
\caption{Gamma-ray spectra, calculated per solid angle, from a proton jet, with a Gaussian intensity
profile (FWHM $3^{\circ}$) centered on the jet axis, observed from four different angles,
$\theta=0^{\circ}$ (solid) $5^{\circ}$ (dashed), $10^{\circ}$ (dotted), and $20^{\circ}$
(dot-dashed). The protons in the jet are assumed to follow a power-law distribution with index
$s=2.0$ and extending up to $T_{p}=512$ TeV.}
\label{fig:jet_spectra}
\end{figure}

The second example is a fanned proton jet which features a Gaussian intensity profile centered on
the jet axis. With a FWHM of $3^{\circ}$ the opening angle of the jet is about $10^{\circ}$. The
gamma-ray spectrum of the jet is integrated over the intensity profile, which is sampled in
$0^{\circ}\!\!.1\times0^{\circ}\!\!.1$ bins, where each bin is represented by the average of ten
randomly sampled pencil beams pointing within the bin. Protons in the jet are again assumed to
follow a power-law distribution with index $s=2.0$ and extending up to $T_{p}=512$ TeV. The
gamma-ray spectra, calculated per solid angle, observed from four different angles,
$\theta=0^{\circ}$ (head on), $5^{\circ}$, $10^{\circ}$, and $20^{\circ}$ are shown in Figure
\ref{fig:jet_spectra}. As with the pencil beam spectra, the absolute normalization is relative to
the density and distribution of target protons.

When the viewing angle is smaller than the opening angle of the jet the gamma-ray spectrum features
a tail extending up to the highest possible gamma-ray energy, about $10^{6}$ GeV, as can be seen in
Figure \ref{fig:jet_spectra}. The tail is suppressed for larger viewing angles because of the
Gaussian intensity profile. At $\theta=5^{\circ}$ the tail is about four orders of magnitude lower
in flux.

\section{Conclusions and Future Prospects}
The angular distribution of gamma-rays produced by proton-proton interactions have been presented in
parameterized formulae. The formulae were derived from Monte Carlo simulations of the up-to-date
proton-proton interaction model by \citet{article:Kamae_etal:2006} and they facilitate computation
of gamma-ray spectra in cases of anisotropic proton distributions. The formulae incorporate all
important known features of the proton-proton interaction up to about $T_{p}=500$ TeV.

As an example of the application of the formulae, gamma-ray spectra was calculated for different
viewing angles from a pencil beam of protons as well as a proton jet with an Gaussian intensity
profile. The pencil beam example shows very clearly that the gamma-ray spectrum changes
significantly as the observer is moved off the beam. The gamma-ray flux decreases drastically and
the spectrum gets cut off at lower energy.

The jet with a Gaussian intensity profile provides a more realistic example. The flux does not
decrease as drastic, but the spectrum changes significantly. For on-axis observers the spectrum
features a prominent tail which is suppressed as the observer is moved off axis. The peak of the
spectrum is shifted to lower energies as the observer is moved.

Particle acceleration models predict some degree of anisotropy for the highest-energy particles
escaping from the acceleration site. The parameterized model presented here can be used to
calculated the anisotropy in gamma-ray emission for any given anisotropy in the proton distribution.

The implementation of formulae and parameters given in this paper in a C language library will be
made available by the authors\footnote{http://www.slac.stanford.edu/$\sim$niklas/cparamlib}. The
functions implemented in this library can be used to calculate both the parameters in Table
\ref{table:pT_params} for any given $T_{p}$ and the differential cross section in equation
\ref{eq:pT_rep} using those parameters for any given set of $E$ and $p_{t}$.

\acknowledgments
The authors would like to acknowledge valuable discussions with and comments received from
T. Abe, K. Andersson, P. Carlson, J. Chiang, J. Cohen-Tanugi, S. Digel, E. do Couto e Silva, T. Koi,
G. Madejski, T. Mizuno, I. Moskalenko, M. M\"uller, P. Nolan, A. Reimer, O. Reimer, T. Sj\"ostrand,
H. Tajima and L. Wai.

N. Karlsson is grateful for the hospitality extended to him by SLAC and KIPAC and the encouragement
given by R. Blandford, S. Kahn and P. Drell.

This work was supported in part by the U.S. Department of Energy under Grant DE-AC02-76SF00515.

\bibliographystyle{apj}
\bibliography{refs}

\begin{thebibliography}{58}
\expandafter\ifx\csname natexlab\endcsname\relax\def\natexlab#1{#1}\fi

\bibitem[{Abe {et~al.}(1988)}]{article:Abe_etal:1988}
Abe, F. {et~al.} 1988, Phys. Rev. Lett., 61, 1819

\bibitem[{Aharonian(2004)}]{book:Aharonian:2004}
Aharonian, F.~A. 2004, Very High Energy Cosmic Gamma Radiation: A Crucial
  Window on the Extreme Universe (World Scientific Publishing)

\bibitem[{Aharonian {et~al.}(2003)}]{article:Aharonian_etal:2003}
Aharonian, F.~A. {et~al.} 2003, A\&A, 403, L1

\bibitem[{Aharonian
  {et~al.}(2004{\natexlab{a}})}]{article:Aharonian_etal:2004a}
---. 2004{\natexlab{a}}, Nature, 432, 75

\bibitem[{Aharonian
  {et~al.}(2004{\natexlab{b}})}]{article:Aharonian_etal:2004b}
---. 2004{\natexlab{b}}, A\&A, 425, L13

\bibitem[{Aharonian {et~al.}(2005)}]{article:Aharonian_etal:2005}
---. 2005, Science, 307, 1938

\bibitem[{Alner {et~al.}(1987)}]{article:Alner_etal:1987}
Alner, G.~J. {et~al.} 1987, Physics Reports, 154, 247

\bibitem[{Alper {et~al.}(1975)}]{article:Alper_etal:1975b}
Alper, B. {et~al.} 1975, Nucl. Phys., B100, 237

\bibitem[{Banner {et~al.}(1982)}]{article:Banner_etal:1982}
Banner, M. {et~al.} 1982, Phys. Lett., 115B, 59

\bibitem[{Banner {et~al.}(1983)}]{article:Banner_etal:1983}
---. 1983, Phys. Lett., 122B, 322

\bibitem[{Beilicke {et~al.}(2005)}]{inproc:Beilicke_etal:2005}
Beilicke, M. {et~al.} 2005, in Proceedings of the 29th International Cosmic Ray
  Conference, ed. B.~S. Acharya {et~al.}, Pune, India

\bibitem[{Berezhko \& Volk(2000)}]{article:BerezhkoVolk:2000}
Berezhko, E.~G. \& Volk, H.~J. 2000, ApJ, 540, 923

\bibitem[{Bicknell \& Begelman(1996)}]{article:BicknellBegelman:1996}
Bicknell, G.~V. \& Begelman, M.~C. 1996, ApJ, 467, 597

\bibitem[{Blattnig {et~al.}(2000)}]{article:Blattnig_etal:2000}
Blattnig, S.~R. {et~al.} 2000, Phys. Rev. D, 62, 094030

\bibitem[{B\"ottcher \& Reimer(2004)}]{article:BottcherReimer:2004}
B\"ottcher, M. \& Reimer, A. 2004, ApJ, 609, 576

\bibitem[{Crawford {et~al.}(1980)}]{article:Crawford_etal:1980}
Crawford, J.~F. {et~al.} 1980, Phys. Rev., C22, 1184

\bibitem[{Dermer(1986{\natexlab{a}})}]{article:Dermer:1986a}
Dermer, C.~D. 1986{\natexlab{a}}, A\&A, 157, 223

\bibitem[{Dermer(1986{\natexlab{b}})}]{article:Dermer:1986b}
---. 1986{\natexlab{b}}, ApJ, 307, 47

\bibitem[{Enomoto {et~al.}(2002)}]{article:Enomoto_etal:2002}
Enomoto, R. {et~al.} 2002, Nature, 416, 823

\bibitem[{Hagiwara {et~al.}(2002)}]{article:Hagiwara_etal:2002}
Hagiwara, K. {et~al.} 2002, Phys. Rev., D66, 010001

\bibitem[{Halzen(2005)}]{inproc:Halzen:2005}
Halzen, F. 2005, in AIP Conference Proceedings, Vol. 745, 2nd International
  Symposium on High Energy Gamma-Ray Astronomy, ed. F.~A. Aharonian, H.~J.
  V\"olk, \& D.~Horns (New York: AIP), 3--13

\bibitem[{Hartman {et~al.}(1999)}]{article:Hartman_etal:1999}
Hartman, R.~C. {et~al.} 1999, ApJS, 123, 79

\bibitem[{Hayakawa(1969)}]{book:Hayakawa:1969}
Hayakawa, S. 1969, Cosmic Ray Physics (John Wiley \& Sons)

\bibitem[{Hunter {et~al.}(1997)}]{article:Hunter_etal:1997}
Hunter, S.~D. {et~al.} 1997, ApJ, 481, 205

\bibitem[{Iyudin {et~al.}(2005)}]{article:Iyudin_etal:2005}
Iyudin, A.~F. {et~al.} 2005, A\&A, 429, 225

\bibitem[{Kamae {et~al.}(2005)Kamae, Abe, \& Koi}]{article:KamaeAbeKoi:2005}
Kamae, T., Abe, T., \& Koi, T. 2005, ApJ, 620, 244

\bibitem[{Kamae {et~al.}(2006)}]{article:Kamae_etal:2006}
Kamae, T. {et~al.} 2006, ApJ, 647, 692, (erratum 662, 779)

\bibitem[{Katagiri {et~al.}(2005)}]{article:Katagiri_etal:2005}
Katagiri, H. {et~al.} 2005, ApJ, 619, L163

\bibitem[{Koers {et~al.}(2006)Koers, Pe'er, \&
  Wijers}]{article:KoersPeerWijers:2006}
Koers, H. B.~J., Pe'er, A., \& Wijers, R. A. M.~J. 2006, preprint
  (hep-ph/0611219)

\bibitem[{Koyama {et~al.}(1997)}]{article:Koyama_etal:1997}
Koyama, K. {et~al.} 1997, PASJ, 49, L7

\bibitem[{Mori(1997)}]{article:Mori:1997}
Mori, M. 1997, ApJ, 478, 225

\bibitem[{Moskalenko {et~al.}(2007)}]{inproc:Moskalenko_etal:2007}
Moskalenko, I.~V. {et~al.} 2007, in Proceedings of the 30th International
  Cosmic Ray Conference, Merida, Mexico

\bibitem[{M\r{a}rtensson {et~al.}(2000)}]{article:Martensson_etal:2000}
M\r{a}rtensson, J. {et~al.} 2000, Phys. Rev., C62, 014610

\bibitem[{M\"ucke \& Protheroe(2001)}]{article:MuckeProtheroe:2001}
M\"ucke, A. \& Protheroe, R.~J. 2001, Astroparticle Physics, 15, 121

\bibitem[{M\"ucke {et~al.}(2003)}]{article:Mucke_etal:2003}
M\"ucke, A. {et~al.} 2003, Astroparticle Physics, 18, 593

\bibitem[{Murthy \& Wolfendale(1986)}]{book:MurthyWolfendale:1986}
Murthy, P. V.~R. \& Wolfendale, A.~W. 1986, Gamma-Ray Astronomy (Cambridge
  University Press)

\bibitem[{Ong(1998)}]{article:Ong:1998}
Ong, R.~A. 1998, Physics Reports, 305, 93

\bibitem[{Reimer {et~al.}(2004)Reimer, Protheroe, \&
  Donea}]{article:ReimerProtheroeDonea:2004}
Reimer, A., Protheroe, R.~J., \& Donea, A.-C. 2004, A\&A, 419, 89

\bibitem[{Rossi {et~al.}(1975)}]{article:Rossi_etal:1975}
Rossi, A.~M. {et~al.} 1975, Nuclear Physics, B84, 269

\bibitem[{Schlickeiser(2002)}]{book:Schlickeiser:2002}
Schlickeiser, R. 2002, Cosmic Ray Astrophysics (Springer)

\bibitem[{Sch\"onfelder(2001)}]{book:Schonfelder:2001}
Sch\"onfelder, V. 2001, The Universe in Gamma Rays (Springer)

\bibitem[{Schroedter {et~al.}(2005)}]{article:Schroedter_etal:2005}
Schroedter, M. {et~al.} 2005, ApJ, 634, 947

\bibitem[{Sj\"ostrand \& Skands(2004)}]{article:SjostrandSkands:2004}
Sj\"ostrand, T. \& Skands, P.~Z. 2004, J. High Energy Phys., 3, 53

\bibitem[{Sj\"ostrand {et~al.}(2001)}]{article:Sjostrand_etal:2001}
Sj\"ostrand, T. {et~al.} 2001, Comput. Phys. Commun., 135, 238

\bibitem[{Slane {et~al.}(1999)}]{article:Slane_etal:1999}
Slane, P. {et~al.} 1999, ApJ, 525, 357

\bibitem[{Stawarz {et~al.}(2006)}]{article:Stawarz_etal:2006}
Stawarz, L. {et~al.} 2006, MNRAS, 370, 981

\bibitem[{Stecker(1970)}]{article:Stecker:1970}
Stecker, F.~W. 1970, Ap\&SS, 377

\bibitem[{Stecker(1973)}]{article:Stecker:1973}
---. 1973, ApJ, 185, 499

\bibitem[{Stecker(1989)}]{inbook:Stecker:1989}
---. 1989, Cosmic Gamma Rays, Neutrinos and Related Astrophysics, ed. M.~M.
  Shapiro \& J.~P. Wefel (Kluwer Academic Publishers), 85--120

\bibitem[{Stephens \& Badhwar(1981)}]{article:StephensBadhwar:1981}
Stephens, S.~A. \& Badhwar, G.~G. 1981, Ap\&SS, 76, 213

\bibitem[{Strong {et~al.}(2007)Strong, Moskalenko, \&
  Ptuskin}]{article:Strong_etal:2007}
Strong, A.~W., Moskalenko, I.~V., \& Ptuskin, V.~S. 2007, Annu. Rev. Nucl.
  Part. Sci., 57, 285

\bibitem[{Strong {et~al.}(2000)Strong, Moskalenko, \&
  Reimer}]{article:Strong_etal:2000}
Strong, A.~W., Moskalenko, I.~V., \& Reimer, O. 2000, ApJ, 537, 763, (erratum
  541, 1109)

\bibitem[{Strong {et~al.}(2004)Strong, Moskalenko, \&
  Reimer}]{article:Strong_etal:2004}
---. 2004, ApJ, 613, 962

\bibitem[{Strong {et~al.}(1978)}]{article:Strong_etal:1978}
Strong, A.~W. {et~al.} 1978, MNRAS, 182, 751

\bibitem[{Strong {et~al.}(1982)}]{article:Strong_etal:1982}
---. 1982, A\&A, 115, 404

\bibitem[{Tsunemi {et~al.}(2000)}]{article:Tsunemi_etal:2000}
Tsunemi, H. {et~al.} 2000, PASJ, 52, 887

\bibitem[{Uchiyama {et~al.}(2003)Uchiyama, Aharonian, \&
  Takahashi}]{article:UchiyamaAharonianTakahashi:2003}
Uchiyama, Y., Aharonian, F.~A., \& Takahashi, T. 2003, A\&A, 400, 567

\bibitem[{Weekes(2003)}]{inproc:Weekes:2003}
Weekes, T.~C. 2003, in Proceedings of the 28th International Cosmic Ray
  Conference, Tsukuba, Japan, 1--12

\end{thebibliography}

\end{document}